\documentclass[12pt]{iopart}

\usepackage{epsfig,multicol,bbm}
\usepackage{setstack}

\begin{document}

\title{Ambiguities in the Seiberg-Witten map and emergent gravity}

\author{Victor O. Rivelles}
\address{Instituto de F\'{i}sica, Universidade de S\~ao Paulo, C. Postal 66318, 05314-970 S\~ao Paulo, SP, Brazil} \eads{rivelles@fma.if.usp.br}



\begin{abstract}The $\theta$ expansion of the Seiberg-Witten map has ambiguities which can be removed by a gauge transformation and/or a field redefinition. In the context of emergent gravity such a field redefinition changes the emerging metric and requires the presence of non-minimal gravitational couplings. It also requires that a real scalar field becomes a scalar density and allows the introduction of a potential. We also find that the potential can have only one term and that a quartic interaction is not allowed. Even though the metric depends on the ambiguity we show that the dispersion relation does not present any sign of it. A proposal for an exact Seiberg-Witten map is used to derive the full metric going beyond the linearized limit.
\end{abstract}

\noindent{\it Keywords\/}: Noncommutative field theories, emergent gravity

\pacs{11.10.Nx,04.50.-h}

\section{Introduction}

Since the advent of the AdS/CFT correspondence the idea that local symmetries are not fundamental acquired a renewed interest and has become a topic of great importance. On one side of the correspondence we can have an ordinary gauge theory at weak coupling in flat spacetime. As the coupling increases the theory is best described as a string theory in curved spacetime. At strong coupling gravity has become an emergent phenomenon. Situations similar to this can happen in several settings and have been the subject of much attention in recent years (for some review papers see \cite{Seiberg:2006wf,Carroll:2010zza,Yang:2011bd,Carlip:2012wa}). Usually the relation between the original theory without gravity and the theory with gravity is very cumbersome so it is desirable to have some situations where this relation can be as simple as possible. Sometime ago this was found in the context of noncommutative (NC) theories \cite{Rivelles:2002ez} where the emerging gravitational field was expressed explicitly in terms of the original matter and gauge fields. 

One of most important properties of noncommutative theories induced by the Moyal product in flat spacetime is the fact that translations in noncommutative directions are equivalent to gauge transformations \cite{Gross:2000ph}. It has a feeling of general relativity so it is natural to look for a connection between gravity and NC theories. This was achieved in \cite{Rivelles:2002ez} where it was shown that after the Seiberg-Witten (SW) map \cite{Seiberg:1999vs} gravity emerges from the NC theory. The effect of the gauge field and the NC parameter $\theta^{\alpha\beta}$ on matter fields induces interactions similar to those produced by gravity. The emergence of gravity through a NC theory lead to a series of applications ranging from matrix models to holography \cite{Yang:2006dk,Yang:2008fb,Yang:2010kj,Steinacker:2010rh,Heckman:2011qu,Lee:2012px,Lee:2012rb,Lee:2012ju}.

Usually a NC field theory is formulated in a NC spacetime by replacing the ordinary product of the fields by the Moyal product. The NC fields are then defined in commutative spacetime but have non conventional properties like for instance self interactions of the gauge field in an Abelian gauge theory. Through the Seiberg-Witten map we can work with conventional fields defined on commutative spacetime at the expenses of introducing a large number of interaction terms. The NC effects appear for instance in modified dispersion relations for photons which no longer move with the velocity of light \cite{Guralnik:2001ax}. In this setting it was found that matter fields interacting with Abelian gauge fields to lowest order in $\theta$ have couplings which are essentially gravitational couplings \cite{Rivelles:2002ez}. These results have been extended to all orders in the NC parameter and at the full non-linear level in the limit of slowly varying gauge fields \cite{Yang:2004vd,Banerjee:2004rs}. 

It is well known that the $\theta$ expansion of the SW map has ambiguities. For instance, for the gauge field the most general expression for the SW map to first order in $\theta^{\mu\nu}$ is given by 
\begin{eqnarray} \label{SWA}
	\hat{A}_\mu = A_\mu - \frac{1}{2} \theta^{\alpha\beta} A_\alpha ( \partial_\beta A_\mu + F_{\beta\mu} ) + \alpha \, \partial_\mu \theta F,
\end{eqnarray}
where $\theta F$ stands for $\theta^{\alpha\beta} F_{\alpha\beta}$. The ambiguity is parametrized by a real constant $\alpha$ and can be eliminated by a gauge transformation on $A_\mu$ with gauge parameter $\Lambda = - \alpha \,\, \theta F$. For a real scalar field we have 
\begin{eqnarray} \label{SWphi}
	\hat{\phi} = \phi - \theta^{\alpha\beta} A_\alpha \partial_\beta \phi + \alpha \, \theta F \, \phi,
\end{eqnarray}
where the ambiguity is again parametrized by a real constant $\alpha$ and it can removed by a field redefinition of the scalar field $\phi  =  \phi^\prime - \alpha \,\, \theta F \,\, \phi^\prime$. A field redefinition changes the action but should not generate any change in the physics. So any physical process evaluated before or after the field redefinition should give the same result. On the other side we know that the effect of the NC gauge field on the scalar field is equivalent to that of emergent gravity \cite{Rivelles:2002ez}. Since gravity is sensitive to any modification that is made in the action we could expect that an ambiguity in the SW map could cause a physical effect in the gravitational context. 

As we shall see, the ambiguity gives rise to a different geometry when compared with the geometry in the absence of the ambiguity. This is discussed in Section \ref{s2}. Firstly, to have emergent gravity we find that the scalar field must be promoted to a scalar density field. We also find that a non-minimal gravitational coupling to the scalar density field is required but a conformal coupling is excluded. We also find that even in the absence of any ambiguity gravity can emerge if the scalar field has density weight $-1/4$. This is the density weight a scalar field would have in order to have conformal symmetry \cite{Sigal:1974,Weinberg:2010fx} but again the conformal coupling is not allowed. In Section \ref{s3} we will see that  we can now include self interactions of the scalar field a situation which was not allowed previously in the absence of ambiguities. We also find that emergent gravity allows only one term in the potential and that the interaction $\phi^4$ is not allowed and again conformal symmetry is prohibited. It is remarkable that the gravity side is sensitive to the presence of noncommutativity and does not allow a conformal theory to emerge. 

In Section \ref{DR} we analyze the dispersion relation for massive and massless particles both in the NC theory and in the linearized gravitational background. We find that the dispersion relation does not depend on the ambiguities and indeed they have the same form in both contexts as expected. Finally, in Section \ref{NL}, we reconsider a proposal for an exact SW map to find the full metric beyond the linearized approximation. When the density weight is different from $-1/4$ we find that $\det g=-1$ so that we get was is called a unimodular gravity. This kind of gravity theory is invariant under volume preserving diffeomorphisms instead of full diffeomorphisms. This sort of gravity theory was found earlier \cite{Harikumar:2006xf} when gravity was extended to NC curved spaces with either the Moyal product or the Kontsevich product. Finally, in the last section we present our conclusions. 

\section{Emergent Gravity in the Presence of Ambiguities}\label{s2}

The action for the NC scalar field $\hat{\phi}$ in the adjoint representation of U(1) without self-interactions in Minkowski spacetime is 
\begin{eqnarray}\label{2.1}
	\hat{S}_0 = \frac{1}{2} \int d^4x \, \hat{D}^\mu \hat{\phi} \star \hat{D}_\mu \hat{\phi},
\end{eqnarray}
where $\hat{D}_\mu \hat{\phi} = \partial_\mu \hat{\phi} - i [ \hat{A}_\mu, \hat{\phi} ]_*$. Applying the SW map (\ref{SWA}) and (\ref{SWphi}) and keeping only first order terms in $\theta$ we get 
\begin{eqnarray}\label{2.2}
\fl 	\hat{S}_0 = \frac{1}{2} \int d^4x \left[ (1 + 2 \alpha \, \theta F) \partial^\mu {\phi} \partial_\mu {\phi} - \alpha {\phi}^2 \opensquare \theta F  - 2 \theta^{\mu\alpha} {F_\alpha}^\nu ( \partial_\mu {\phi} \partial_\nu {\phi} - \frac{1}{4} \eta_{\mu\nu} \partial^\lambda {\phi} \partial_\lambda \phi) \right]. 
\end{eqnarray}
Notice that the term inside the parenthesis is traceless. 

Consider now the action for a scalar density field $\phi$ with weight $-\omega$ (the weight of $\sqrt{-g}$ is $1$) without self-interactions in a gravitational background non-minimally coupled to the curvature scalar
\begin{eqnarray}\label{2.3}
	S^g_0 = \frac{1}{2} \int d^4x \, (\sqrt{-g})^{2\omega+1} \, g^{\mu\nu} D_\mu \phi \, D_\nu\phi + \frac{1}{2} \mu \int d^4x \, (\sqrt{-g})^{2\omega+1} R \, \phi^2, 
\end{eqnarray}
where $\mu$ is the coupling constant and $D_\mu \phi = \partial_\mu + \omega \Gamma^\nu_{\mu\nu} \phi$ is the covariant derivative of a density scalar of weight $-\omega$. Taking the linearized limit $g_{\mu\nu} = \eta_{\mu\nu} + h_{\mu\nu} + \eta_{\mu\nu} h$, where $h_{\mu\nu}$ is traceless, we get 
\begin{eqnarray} \label{gravaction}
\fl 	S^g_0 = \frac{1}{2} \int d^4x \left[ \left(1 + (1+4\omega) h \right) \partial^\mu \phi \partial_\mu \phi - h^{\mu\nu} \partial_\mu\phi \partial_\nu\phi + ( 3\mu - 2\omega ) \phi^2 \opensquare h  - \mu \, \phi^2  \partial^\mu \partial^\nu h_{\mu\nu} \right]. 	
\end{eqnarray}
After identifying the coefficients of $\partial\phi \, \partial\phi$ and $\phi^2$ terms in (\ref{2.2}) with those of (\ref{gravaction}) we get for $\omega \not= -1/4$ that 
\begin{eqnarray}
 h^{\mu\nu} &= \theta^{\mu\alpha}{F_\alpha}^\nu	+ \theta^{\nu\alpha}{F_\alpha}^\mu + \frac{1}{2} \eta^{\mu\nu} \theta F, \label{2.5}\\
h &= -\frac{\mu}{1+6\mu} \theta F,  \label{2.6} \\ 
\mu &= - \frac{1}{6 + \frac{1+4\omega}{2\alpha}}, \label{2.7} 
\end{eqnarray}
for $\alpha$ arbitrary. The non-minimal coupling and the density weight are required by the terms of the form $\phi^2 \opensquare \theta F$. If they are not present there are no consistent solution except when $\alpha=0$. Notice that only the combination $2\alpha/(1+4\omega)$ appears in the solution. Also $\omega \not= -1/4$ implies that the conformal coupling $\mu = -1/6$ is not allowed. When the ambiguity is not present we get $\mu=0$ and $h=0$ recovering the results found in \cite{Rivelles:2002ez}. 

For $\omega = -1/4$ we find that $\alpha = 0$ while $\mu$ remains arbitrary but different from $-1/6$. The linearized metric still is given by (\ref{2.5}) and the trace of metric is still (\ref{2.6}) but now (\ref{2.7}) no longer holds.  

We then find that the geometry depends on the ambiguity in the SW map through the trace of the metric (\ref{2.6}) and (\ref{2.7}) when $\omega \not= -1/4$. However, when we consider the dispersion relation for a particle in this background, in Section \ref{DR}, we will show that no ambiguity dependence is found. 

It is also worth to remark that in the absence of ambiguities there is a new situation that was not detected in \cite{Rivelles:2002ez}. It corresponds to a scalar density field with $\omega=-1/4$ which, as will see in Section \ref{NL}, has a full non-linear completion. 

\section{Including Self Interactions}\label{s3}

As remarked before, in the absence of ambiguities no self interactions were allowed. The reason for that was the imposition that the scalar field in the emergent case was a true scalar field. By relaxing this condition and allowing it to be a density instead of a scalar will allow the presence of interactions. So consider a potential for the NC field $\hat{\phi}$ which is polynomial 
\begin{eqnarray}
	\hat{S}_i = \int d^4 x \, \hat{V}(\hat{\phi}), \quad \hat{V}(\hat{\phi}) = \sum_{n>1} \frac{1}{n} V^{(n)} \hat{\phi}^n.
\end{eqnarray}
After the SW map we get 
\begin{eqnarray} \label{3.2}
	\hat{S}_i = \int d^4 x \sum_{n>1} \left[ \left(1 - \frac{1}{2}(1  - 2n \alpha ) \theta F \right) \frac{1}{n} V^{(n)} \phi^n \right]. 
\end{eqnarray}
In the gravity side we have
\begin{eqnarray}
	S_i^g = \int d^4x \sqrt{-g} \,\, \hat{V}\left( (\sqrt{-g})^\omega \phi \right), 
\end{eqnarray}
which at the linearized level yields
\begin{eqnarray}
	S_i^g = \int d^4x \sum_{n>1} \left[ 1 + 2(1+n\omega) h \right] \frac{1}{n} V^{(n)} \phi^n.
\end{eqnarray}

Identifying the terms with $\phi^n$ in both actions we get an equation which can be solved for the density  weight as 
\begin{eqnarray}\label{omega}
	\omega = -\frac{1}{4} \left[ 1 + 2\alpha(4-n) \right], \quad \omega \not= -\frac{1}{4}
\end{eqnarray}
or written like $n-4= (1+4\omega)/2\alpha$ showing again that only the combination $2\alpha/(1+4\omega)$ is relevant and it is now fixed by the self interaction of $\phi$. Notice that the mass term in the potential contributes with terms of the form $\theta F$ while the contribution from the previous section in $\phi^2$ have the form $\opensquare \theta F$. Combining with (\ref{2.7}) we get 
\begin{eqnarray} \label{3.6}
	h = \frac{1}{n-4} \theta F, \quad \mu = - \frac{1}{n+2}, 
\end{eqnarray}
so that the ambiguity only appears in the density weight (\ref{omega}). 

The important information in (\ref{omega}) is that only one monomial is allowed in the potential since a choice of $n$ fixes the value of the density weight. So emergent gravity allows only one type of self interaction. Notice also that the renormalizable $\phi^4$ interaction is not allowed since it implies that $\omega=-1/4$. This is also seen in (\ref{3.6}). It is also interesting to note that 
\begin{eqnarray} \label{3.7}
	\frac{1}{n-4} = - \frac{\mu}{1+6\mu},
\end{eqnarray}
so that the conformal coupling is also not allowed and is tightly related to the absence of the $\phi^4$ interaction. The situation when both $\alpha$ and $\omega$ vanish, which goes back to the case studied in \cite{Rivelles:2002ez}, is not allowed. 

It should also be remarked that adding non minimal couplings to the potential in the form
\begin{eqnarray}
	S^g_{nm} = \int d^4 x \, R \, \hat{V}(\phi),
\end{eqnarray}
does not help in relaxing (\ref{omega}) since it produces equations for $\opensquare \theta F$ and not for $\theta F$. 

When $\omega = -1/4$ we found before that $\alpha=0$ and remarkably we get the same solution (\ref{3.6}) and the relation (\ref{3.7}) while (\ref{omega}) is no longer true. Again the conformal coupling is not allowed as well as the $\phi^4$ interaction. 

The effect of the self interaction of the scalar field is to fix its density weight. The renormalizable case $n=4$ is excluded as well as the conformal coupling to gravity. The only dependence on the ambiguity resides in the density weight when $\omega \not= -1/4$ since the metric is no longer $\alpha$ dependent. The interaction has washed out the ambiguity in the metric and its only left over is in the density weight. As we shall see, the dispersion relation does not depend on the density weight so we do not expect any ambiguity in this case. 

To summarize we found that the gravitational background is given by 
\begin{eqnarray}
	h^{\mu\nu} &= \theta^{\mu\alpha}{F_\alpha}^\nu	+ \theta^{\nu\alpha}{F_\alpha}^\mu + \frac{1}{2} \eta^{\mu\nu} \theta F, \\
h &= -\frac{\mu}{1+6\mu} \theta F,
\end{eqnarray}
where, for $\omega \not= -1/4$
\begin{eqnarray}
	\mu = \left\{
	\begin{array}{l l}
	- {1}/({6 + \frac{1+4\omega}{2\alpha}}), \quad \omega \quad \mbox{arbitrary} &\mbox{if} \quad V=0, \\
	- {1}/({n+2}), \qquad \,\,\, \omega = -\left[ 1 + 2\alpha(4-n) \right]/4 \qquad &\mbox{if}\quad V\not=0,  
	\end{array} \right.
\end{eqnarray}
while for $\omega = -1/4$ we have $\alpha=0$ and 
\begin{eqnarray}
	\mu = \left\{
	\begin{array}{l l}
	\mbox{arbitrary} &\mbox{if} \quad V=0, \\
	-1/(n+2) \qquad &\mbox{if} \quad V\not=0. 
	\end{array} \right.
\end{eqnarray}
When $V\not=0$ we must have $n\not=4$. 

The linearized Ricci scalar corresponding to this background is
\begin{eqnarray}
	R = \frac{1}{2(1+6\mu)} \opensquare \theta F = \frac{1}{2}\frac{n+2}{n-4} \opensquare \theta F,
\end{eqnarray}
where the last equality holds only in $V\not=0$. For $\omega\not=-1/4$ and $V=0$ it depends on the ambiguity while for $\omega\not=-1/4$ and $V\not=0$ or $\omega=-1/4$ there is no ambiguity. 

\section{Dispersion Relations \label{DR}} 

It is seen then that not only the background but also the Ricci scalar depends on the ambiguity if $\omega\not=-1/4$ while in the NC field theory we expect that any physical process be independent of $\alpha$ even though it appears explicitly in the action (\ref{2.2}) and (\ref{3.2}). To compare both sides and see how they depend on the ambiguity let us consider plane waves in the NC gauge theory. Upon quantization the dispersion relation of the plane waves will give the velocity of the particle associated to the scalar field. We can then look for the velocity of these particles in the gravitational background and derive its gravitational dispersion relation. 

So let us compute the dispersion relation in the gauge theory side for the massive case. Let us assume that the field strength in (\ref{2.2}) is constant and look for plane wave solutions. We find that 
\begin{eqnarray}
	\left[ 1 - \left( \frac{1}{2} - 2\alpha \right) \theta F \right] ( k^2 - m^2 ) - 2 \theta^{\alpha\beta} {F_{\beta}}^\mu k_\mu k_\alpha = 0. 
\end{eqnarray}
To find how the energy depends on the velocity we multiply this equation by $\left[ 1 + \left( {1}/{2} - 2\alpha \right) \theta F \right]$ so that 
\begin{eqnarray} \label{4.2}
	k^2 - m^2  - 2 \theta^{\alpha\beta} {F_{\beta}}^\mu k_\mu k_\alpha = 0,
\end{eqnarray}
and all $\alpha$ dependence is gone away. Then to lowest order in $\theta$ the dispersion relation is not affected by the ambiguity in the SW map as expected. 

In the gravity side we can derive the dispersion relation from $g_{\mu\nu} P^\mu P^\nu - m^2=0$. We then find that 
\begin{eqnarray}
	\left[ 1 - \left( \frac{1}{2} - \frac{\mu}{1+6\mu} \right) \theta F \right] P^2 - 2 \theta^{\alpha\beta} {F_{\beta}}^\mu P_\mu P_\alpha - m^2 = 0,
\end{eqnarray}
where $P^2 = \eta^{\mu\nu}P_\mu P_\nu$. Multiplying by $\left[ 1 + \left( {1}/{2} - {\mu}/({1+6\mu}) \right) \theta F \right]$ we get 
\begin{eqnarray} \label{4.4} 
	P^2 - 2 \theta^{\alpha\beta} {F_{\beta}}^\mu P_\mu P_\alpha - m^2 \left[ 1 + \left( \frac{1}{2} - \frac{\mu}{1+6\mu} \right) \theta F \right] = 0.  
\end{eqnarray}
For the massless case, when $\mu$ depends explicitly on the ambiguity, the dispersion relation has no  dependence on $\alpha$.  For the massive case we have $n=2$ so $\mu=-1/4$ and there is no dependence on $\alpha$ in the dispersion relation. Only the density weight depends on $\alpha$. Then in both cases the ambiguity does not contribute to the dispersion relation and in fact the gravitational dispersion relation coincides with the gauge theory one since the $\theta F$ contribution in the mass term of (\ref{4.4}) vanishes for $n=2$ and we get an equation with the same form as (\ref{4.2}). 

\section{Going to Higher Orders} \label{NL}

Up to now we have been working with the SW map to first order in $\theta$ and in the linearized approximation in the gravity side so that the linearized metric is also first order in $\theta$. To go to the nonlinear level in the gravity side we need higher order $\theta$ terms in the SW map. 
There are some proposals for the SW map which go beyond first order. An exact SW map was obtained in the limit of slowly varying fields by a clever coordinate transformation involving the gauge field  \cite{Yang:2004vd}. It can also be applied to the scalar field case \cite{Banerjee:2004rs} and the resulting NC action after the SW map is given by 
\begin{eqnarray}\label{yang}
\hat{S}_0 = \frac{1}{2} \int d^4x \,\, \sqrt{\det (1 + F\theta)} \left( \frac{1}{1+ F\theta} \frac{1}{1 + \theta F} \right)^{\mu\nu} \partial_\mu \phi  \partial_\nu \phi,	
\end{eqnarray}
where a matrix notation was adopted so that $(1 + F \theta )_{\mu\nu}$ means the matrix $\eta_{\mu\nu} + F_\mu^\lambda \theta_{\lambda\nu}$. When expanded in $\theta$ we find (\ref{2.2}) to lowest order with $\alpha=0$ so that there is no ambiguity. Now we can compare (\ref{yang}) with the emergent gravity action (\ref{2.3}) in the limit of slowly varying field to get 
\begin{eqnarray}
	(\sqrt{-g})^{-2\omega-1} g_{\mu\nu} = \frac{1}{\sqrt{\det(1 + F\theta)}} \left( (1+F\theta)^T(1+F\theta)\right)_{\mu\nu},
\end{eqnarray}
showing explicitly that the metric is symmetric. We then find that if $\omega \not= -1/4$ the metric is
\begin{eqnarray} \label{5.3}
	g_{\mu\nu} = \frac{1}{\sqrt{\det(1 + F\theta)}} \left( (1+F\theta)^T(1+F\theta)\right)_{\mu\nu}, \qquad \omega \not= -\frac{1}{4}, 
\end{eqnarray}
with $\det{g}=-1$. From the results in Section \ref{s2} we find that if $\alpha=0$ then $\mu=0$ so that the trace of linearized metric vanishes $h=0$. (Notice however that the trace of (\ref{5.3}) is non-vanishing and only its linearized value gives zero.) Now, if $\omega = -1/4$ we find
\begin{eqnarray}
\fl	g_{\mu\nu} = \left(\sqrt{\det(1 + F\theta)}\right)^{-(1+4\mu)/(1+6\mu)} \left( (1+F\theta)^T(1+F\theta)\right)_{\mu\nu}, \qquad \omega = -\frac{1}{4},
\end{eqnarray}
and $\det g = \det(1+F\theta)^{4\mu/(1+6\mu)}$. Again, from the results of Section \ref{s2} we find that $\mu$ is arbitrary and different from $-1/6$ and the trace of the linearized metric is given by (\ref{2.6}). 

Since we are in a broader context our results differ from \cite{Banerjee:2004rs} where $\phi$ was regarded as a true scalar field. To take into account the usual $\sqrt{-g}$ contribution to the action of the scalar field a dilaton was introduced to give the required power of $\det(1 + F\theta)$. If we take $\omega=0$ in (\ref{5.3}) we get $\det g = -1$ and we reproduce the result coming from the exact SW map.

It is also worth remarking that when $\omega \not= -1/4$ we have $\det g=-1$ which is characteristic of a class of gravity theories dubbed unimodular gravity theories which are invariant by volume preserving diffeomorphisms (for a review see (\cite{Alvarez:2005iy}). Gravity theories with volume preserving diffeomorphisms in the NC setting were derived and analyzed in \cite{Harikumar:2006xf}. 

When going to higher orders in the SW map we realize that up to second order the $\theta^2$ contributions to the action (\ref{2.2}) can still be put in the form $g^{\mu\nu} \partial_\mu \phi \partial_\nu \phi$ with first order derivatives acting on the scalar field. However at higher orders this is no longer true since the Moyal product in (\ref{2.1}) gives rise to terms with more than one derivative acting on the scalar field and also contributions which go beyond the limit of slowly varying fields. This means that the usual coupling to gravity is no longer valid. In fact it points out in the direction that some version of noncommutative gravity is required maybe along the lines of \cite{Harikumar:2006xf} where we proposed extensions involving the Moyal product and the Kontsevich product and it was found that both required $\det g =-1$. An explicit contribution to the SW map to order $\theta^3$ for the action of the scalar field was computed in \cite{Stern:2009vx} and we are investigating its compatibility with the noncommutative gravity theories proposed in \cite{Harikumar:2006xf}.

\section{Conclusions}

We have analyzed the consequences of the ambiguity of the SW map in the emergent gravity context. We considered the case of a scalar field in the adjoint representation of $U(1)$ in the NC theory and found that a gravitational interpretation requires that the scalar field turns into a scalar density field. It also requires that a non-minimal gravitational coupling to the scalar field is turned on. When the density weight is $-1/4$ and the scalar density field is indeed conformal invariant the ambiguity is no longer required. In general a potential for the scalar field can be added but only if it has just one term. It is interesting that a quartic interaction which would lead to a conformal invariant theory is not allowed. It seems to exist a clash between noncommutativity and conformal symmetry. The emergent metric depends on the ambiguity so it is necessary to verify whether the resulting physical effects also depend on the ambiguity. In the NC theory the ambiguity can not affect the physics. This is confirmed by computing the dispersion relation for plane waves. In the gravity side we also computed the dispersion relation for particles and even though the metric depends on the ambiguity the dispersion relation is ambiguity free. 

The extension of these results to all orders in the NC parameter and also to the full non-linear level in the gravity side can be done in the case of slowly varying gauge field and in the absence of ambiguities. Further work is required to go beyond this limit since it will also require the knowledge of some form of noncommutative gravity theory or even generalized geometry present in NC theories \cite{Jurco:2013upa}.

It is also well known that supersymmetric NC theories have important properties. It was expected that supersymmetry could remove the mixing of UV and IR divergences characteristic of NC field theories \cite{Minwalla:1999px} but this happens only in the absence of gauge fields \cite{Girotti:2000gc}. After the use of the SW map it is also known that the supersymmetry algebra presents serious troubles \cite{Mikulovic:2003sq} and we expect that the inclusion of the ambiguity may help in the understanding of theses difficulties.

\ack

V.O.R. wants to acknowledge the participation of Ciro Yajima at the initial stages of this work and comments by H. S. Yang. This work was supported by CNPq grant 304116/2010-6 and FAPESP grant 2008/05343-5.

\section*{References}

\begin{thebibliography}{999}

\bibitem{Seiberg:2006wf}
N.~Seiberg, {\it {Emergent spacetime}},
  {http://xxx.lanl.gov/abs/hep-th/0601234}{{\tt hep-th/0601234}}.

\bibitem{Carroll:2010zza}
R.~Carroll, {\it {On the emergence theme of physics, World Scientific
  Publishing (2010)}}, .

\bibitem{Yang:2011bd}
H.~S. Yang, {\it {Towards A Background Independent Quantum Gravity}},  {\em
  J.Phys.Conf.Ser.} {\bf 343} (2012) 012132,
  [{http://xxx.lanl.gov/abs/1111.0015}{{\tt arXiv:1111.0015}}].

\bibitem{Carlip:2012wa}
S.~Carlip, {\it {Challenges for Emergent Gravity}},
  {http://xxx.lanl.gov/abs/1207.2504}{{\tt arXiv:1207.2504}}.

\bibitem{Rivelles:2002ez}
V.~O. Rivelles, {\it {Noncommutative field theories and gravity}},  {\em
  Phys.Lett.} {\bf B558} (2003) 191--196,
  [{http://xxx.lanl.gov/abs/hep-th/0212262}{{\tt hep-th/0212262}}].

\bibitem{Gross:2000ph}
D.~J. Gross and N.~A. Nekrasov, {\it {Dynamics of strings in noncommutative
  gauge theory}},  {\em JHEP} {\bf 0010} (2000) 021,
  [{http://xxx.lanl.gov/abs/hep-th/0007204}{{\tt hep-th/0007204}}].

\bibitem{Seiberg:1999vs}
N.~Seiberg and E.~Witten, {\it {String theory and noncommutative geometry}},
  {\em JHEP} {\bf 9909} (1999) 032,
  [{http://xxx.lanl.gov/abs/hep-th/9908142}{{\tt hep-th/9908142}}].

\bibitem{Yang:2006dk}
H.~S. Yang, {\it {Emergent Gravity from Noncommutative Spacetime}},  {\em
  Int.J.Mod.Phys.} {\bf A24} (2009) 4473--4517,
  [{http://xxx.lanl.gov/abs/hep-th/0611174}{{\tt hep-th/0611174}}].

\bibitem{Yang:2008fb}
H.~S. Yang, {\it {Emergent Spacetime and The Origin of Gravity}},  {\em JHEP}
  {\bf 0905} (2009) 012, [{http://xxx.lanl.gov/abs/0809.4728}{{\tt
  arXiv:0809.4728}}].

\bibitem{Yang:2010kj}
H.~S. Yang, {\it {Emergent Geometry and Quantum Gravity}},  {\em
  Mod.Phys.Lett.} {\bf A25} (2010) 2381--2397,
  [{http://xxx.lanl.gov/abs/1007.1795}{{\tt arXiv:1007.1795}}].

\bibitem{Steinacker:2010rh}
H.~Steinacker, {\it {Emergent Geometry and Gravity from Matrix Models: an
  Introduction}},  {\em Class.Quant.Grav.} {\bf 27} (2010) 133001,
  [{http://xxx.lanl.gov/abs/1003.4134}{{\tt arXiv:1003.4134}}].

\bibitem{Heckman:2011qu}
J.~J. Heckman and H.~Verlinde, {\it {Instantons, Twistors, and Emergent
  Gravity}},  {http://xxx.lanl.gov/abs/1112.5210}{{\tt arXiv:1112.5210}}.

\bibitem{Lee:2012px}
S.~Lee, R.~Roychowdhury, and H.~S. Yang, {\it {Notes on Emergent Gravity}},
  {\em JHEP} {\bf 1209} (2012) 030,
  [{http://xxx.lanl.gov/abs/1206.0678}{{\tt arXiv:1206.0678}}].

\bibitem{Lee:2012rb}
S.~Lee, R.~Roychowdhury, and H.~S. Yang, {\it {Test of Emergent Gravity}},
  {http://xxx.lanl.gov/abs/1211.0207}{{\tt arXiv:1211.0207}}.

\bibitem{Lee:2012ju}
S.~Lee, R.~Roychowdhury, and H.~S. Yang, {\it {Topology Change of Spacetime and
  Resolution of Spacetime Singularity in Emergent Gravity}},
  {http://xxx.lanl.gov/abs/1212.3000}{{\tt arXiv:1212.3000}}.

\bibitem{Guralnik:2001ax}
Z.~Guralnik, R.~Jackiw, S.~Pi, and A.~Polychronakos, {\it {Testing
  noncommutative QED, constructing noncommutative MHD}},  {\em Phys.Lett.} {\bf
  B517} (2001) 450--456, [{http://xxx.lanl.gov/abs/hep-th/0106044}{{\tt
  hep-th/0106044}}].

\bibitem{Yang:2004vd}
H.~S. Yang, {\it {Exact Seiberg-Witten map and induced gravity from
  noncommutativity}},  {\em Mod.Phys.Lett.} {\bf A21} (2006) 2637--2647,
  [{http://xxx.lanl.gov/abs/hep-th/0402002}{{\tt hep-th/0402002}}].

\bibitem{Banerjee:2004rs}
R.~Banerjee and H.~S. Yang, {\it {Exact Seiberg-Witten map, induced gravity and
  topological invariants in noncommutative field theories}},  {\em Nucl.Phys.}
  {\bf B708} (2005) 434--450,
  [{http://xxx.lanl.gov/abs/hep-th/0404064}{{\tt hep-th/0404064}}].

\bibitem{Sigal:1974}
R.~F. Sigal, {\it {Conformal Invariance and the Six-Dimensional Formalism}},
  {\em Int.J.Theor.Phys.} {\bf 11} (1974) 45--68.

\bibitem{Weinberg:2010fx}
S.~Weinberg, {\it {Six-dimensional Methods for Four-dimensional Conformal Field
  Theories}},  {\em Phys.Rev.} {\bf D82} (2010) 045031,
  [{http://xxx.lanl.gov/abs/1006.3480}{{\tt arXiv:1006.3480}}].

\bibitem{Harikumar:2006xf}
E.~Harikumar and V.~O. Rivelles, {\it {Noncommutative Gravity}},  {\em
  Class.Quant.Grav.} {\bf 23} (2006) 7551--7560,
  [{http://xxx.lanl.gov/abs/hep-th/0607115}{{\tt hep-th/0607115}}].

\bibitem{Alvarez:2005iy}
E.~Alvarez, {\it {Can one tell Einstein's unimodular theory from Einstein's
  general relativity?}},  {\em JHEP} {\bf 0503} (2005) 002,
  [{http://xxx.lanl.gov/abs/hep-th/0501146}{{\tt hep-th/0501146}}].

\bibitem{Stern:2009vx}
A.~Stern, {\it {Remarks on an Exact Seiberg-Witten map}},  {\em Phys.Rev.} {\bf
  D80} (2009) 067703, [{http://xxx.lanl.gov/abs/0907.4532}{{\tt
  arXiv:0907.4532}}].

\bibitem{Jurco:2013upa}
B.~Jurco, P.~Schupp, and J.~Vysoky, {\it {On the Generalized Geometry Origin of
  Noncommutative Gauge Theory}},  {http://xxx.lanl.gov/abs/1303.6096}{{\tt
  arXiv:1303.6096}}.

\bibitem{Minwalla:1999px}
S.~Minwalla, M.~Van~Raamsdonk, and N.~Seiberg, {\it {Noncommutative
  perturbative dynamics}},  {\em JHEP} {\bf 0002} (2000) 020,
  [{http://xxx.lanl.gov/abs/hep-th/9912072}{{\tt hep-th/9912072}}].

\bibitem{Girotti:2000gc}
H.~Girotti, M.~Gomes, V.~O. Rivelles, and A.~da~Silva, {\it {A Consistent
  noncommutative field theory: The Wess-Zumino model}},  {\em Nucl.Phys.} {\bf
  B587} (2000) 299--310, [{http://xxx.lanl.gov/abs/hep-th/0005272}{{\tt
  hep-th/0005272}}].

\bibitem{Mikulovic:2003sq}
D.~Mikulovic, {\it {Seiberg-Witten map for superfields on canonically deformed
  N = 1, d = 4 superspace}},  {\em JHEP} {\bf 0401} (2004) 063,
  [{http://xxx.lanl.gov/abs/hep-th/0310065}{{\tt hep-th/0310065}}].


\end{thebibliography}

\end{document}